\documentclass[twocolumn,secnumarabic,amssymb, nobibnotes, superscriptaddress, nofootinbib,aps,prd]{revtex4-2}
\usepackage[colorlinks,citecolor=blue,linkcolor=blue,anchorcolor=blue,filecolor=blue, urlcolor=blue]{hyperref}
\usepackage{natbib}

\usepackage{amsmath}
\usepackage{adjustbox}
\usepackage{graphicx}
\usepackage{booktabs}       
\usepackage{amsfonts}      
\usepackage{nicefrac}       
\usepackage{microtype}
\usepackage{cleveref}
\usepackage{xcolor}
\usepackage{multirow}
\usepackage{makecell}
\usepackage{array}
\usepackage{orcidlink}
\usepackage{float}
\usepackage{lineno}
\usepackage{slashed}
\usepackage{bm}
\definecolor{dora}{RGB}{255,150,0}
\usepackage{xcolor}

\usepackage[normalem]{ulem}
\usepackage{caption}
\DeclareCaptionJustification{fulljustify}{\leftskip=0pt \rightskip=0pt \parfillskip=0pt plus 1fil}
\usepackage{tikz}
\usepackage{tikz-feynman}
\usepackage{caption}
\usepackage{subcaption}

\begin{document}

\author{Adamu Issifu \orcidlink{0000-0002-2843-835X}} 
\email{ai@academico.ufpb.br}
\affiliation{CFisUC, Department of Physics, University of Coimbra, 3004-516 Coimbra, Portugal}
\affiliation{Departamento de F\'isica, Instituto Tecnol\'ogico de Aeron\'autica, DCTA, 12228-900, S\~ao Jos\'e dos Campos, SP, Brazil} 
\affiliation{Laborat\'orio de Computa\c c\~ao Cient\'ifica Avan\c cada e Modelamento (Lab-CCAM), Brazil}

\author{Constan\c ca Provid\^encia \orcidlink{0000-0001-6464-8023}}
\email{cp@uc.pt}
\affiliation{CFisUC, Department of Physics, University of Coimbra, 3004-516 Coimbra, Portugal}

\author{Tobias Frederico \orcidlink{0000-0002-5497-5490}} 
\email{tobias@ita.br}
\affiliation{Departamento de F\'isica, Instituto Tecnol\'ogico de Aeron\'autica, DCTA, 12228-900, S\~ao Jos\'e dos Campos, SP, Brazil}
\affiliation{Laborat\'orio de Computa\c c\~ao Cient\'ifica Avan\c cada e Modelamento (Lab-CCAM), Brazil}

\title{Trace-anomaly decomposition and universal dark matter scaling in compact stars}

\begin{abstract}
We investigate how dark matter (DM) admixture modifies the conformal properties and phase structure of dense neutron-star matter within a self-consistent single-fluid framework, with the global DM fraction $F_\chi=N_\chi/N_B$ fixing the local relation $n_\chi=F_\chi n_B$. We derive an exact decomposition of the total trace anomaly, $\Delta_{\rm tot}$, into microscopic contributions. For collider-motivated Higgs-portal benchmarks, explicit Higgs, vector-mediator, and contact-interaction contributions are negligible, while heavy nonrelativistic DM has an intrinsic trace anomaly close to the nonrelativistic limit, $\Delta_\chi\simeq1/3$. Consequently, the DM rest-mass energy fraction dominates the DM-induced modification of $\Delta_{\rm tot}$, producing a smooth upward shift of up to $\sim0.1$ for $F_\chi\sim0.2\%$. In the pressureless, comoving heavy-WIMP regime, we further identify a universal dark-sector scaling governed by the mass-loading parameter $\lambda=F_\chi m_\chi/m_N$: numerical calculations with different $(F_\chi,m_\chi)$ pairs at fixed $\lambda$ exhibit overlapping trace-anomaly, sound-speed, and mass--radius responses for a given baryonic equation of state. In hybrid stars, DM leaves the coexistence pressure and chemical potential essentially unchanged, whereas first-order hadron--quark deconfinement produces sharp discontinuities in the squared sound speed $c_s^2$, $\Delta_{\rm tot}$, and $\Delta_{\rm tot}-\Delta_B$. The combined softening substantially reduces the maximum stellar mass, placing $F_\chi\sim0.2\%$ in tension with the observed $2\,M_\odot$ neutron stars. These results show that the trace anomaly and its density evolution provide a sensitive diagnostic for distinguishing smooth DM-induced modifications of dense matter from genuine first-order deconfinement.
\end{abstract}

\maketitle

\section{Introduction}
Neutron stars (NSs) are composed of ultra-dense nuclear matter \cite{Burgio:2021vgk}, with typical masses of about $1.4\,M_\odot$ and radii of about $13\,\mathrm{km}$, supported against gravitational collapse by the pressure of strongly interacting, degenerate matter \cite{Lattimer:2000nx,1996csnp.bookGlen}. Recent observations indicate masses approaching $2\,M_\odot$ and radii of about $12\,\mathrm{km}$ \cite{Miller:2021qha,Riley:2021pdl}. At sufficiently high densities, additional degrees of freedom, including hyperons, kaons, and $\Delta$ resonances, may emerge \cite{LATTIMER2006479, Jiang:2012hy, Issifu:2023qyi}, while deconfinement to quark matter may lead to hybrid stars \cite{Glendenning:1992vb, Alford:2013aca, Annala:2019puf, Issifu:2023ovi}. Although such exotic degrees of freedom have not been observationally established, their possible presence remains an open question, motivating theoretical studies constrained by multimessenger observations \cite{Yunes:2022ldq}.

The study of dark matter (DM) in neutron stars (NSs) has attracted growing interest because of its unknown nature and largely unconstrained interactions with baryonic matter (BM) \cite{Bertone:2004pz, Bertone:2018krk}. The lack of a confirmed terrestrial signal \cite{XENON:2025vwd, LZ:2024zvo}, despite strong cosmological \cite{Springel:2005nw, Planck:2015fie} and astrophysical \cite{1980ApJ...238..471R, 1981AJ.....86.1825B, Clowe:2006eq} evidence, motivates NSs as complementary probes. Two main frameworks are commonly employed: a two-fluid approach, where DM and BM obey separate EOSs and interact only gravitationally through the two-fluid TOV equations \cite{Karkevandi:2021ygv, Issifu:2025gsq, Kain:2021hpk}, and a single-fluid approach, where the two sectors are coupled through non-gravitational interactions \cite{Das:2020vng, Gresham:2018rqo, Lopes:2024ixl, Shahrbaf:2025hsw}. In neutron dark decay, a related single-fluid scenario, chemical equilibrium $\mu_n=\mu_\chi$ determines the DM abundance, which can generate substantial DM fractions and soften the EOS, requiring additional repulsive interactions to satisfy massive-NS constraints \cite{Baym:2018ljz, Husain:2022bxl, 2023Univ....9..484Z}. In these scenarios, DM can alter stellar structure and increase compactness, potentially shifting the onset of exotic degrees of freedom, such as hyperons, to lower densities \cite{Issifu:2025jac, Issifu:2024htq}.

 The focus of this manuscript is to quantitatively determine how DM modifies the conformal and phase structure of NSs within a self-consistent single-fluid framework, and to establish whether its thermodynamic signatures can be distinguished from those of genuine hadron--quark deconfinement. Previous work has shown that DM can lower the critical stellar mass for quark-matter formation in a two-fluid, gravitationally coupled framework \cite{Biesdorf:2024dor}, while its impact on conformality has been examined separately in neutron-dark-decay scenarios \cite{Thakur:2024btu} and in interacting DM-admixed matter at fixed entropy per baryon \cite{Das:2026hbu}. These studies establish that DM can modify the high-density thermodynamic response, but the connection between explicit DM--BM interactions, the trace anomaly, and the signatures of deconfinement has not been systematically quantified. 

Here we address this gap using a self-consistent single-fluid framework based on the global DM fraction $F_\chi=N_\chi/N_B$, introduced in \cite{Issifu:2026diw} and subsequently applied in \cite{Issifu:2026ehm}, which, for a thermalized and comoving DM component, imposes the local closure $n_\chi=F_\chi n_B$. Within a Higgs-portal model supplemented by a massive vector mediator, we derive an exact energy-density-weighted decomposition of the normalized trace anomaly introduced in \cite{Fujimoto:2022ohj, Annala:2023cwx}, isolating the contributions of BM, DM, scalar, and vector-interaction sectors and quantifying how DM modifies the approach to conformality. We then extend the analysis to hybrid stars with a genuine first-order hadron--quark transition, using two modified density-dependent quark-mass (MDDQM) model parametrizations, to determine whether DM alters the coexistence conditions or instead leaves the phase boundary essentially unchanged while modifying the thermodynamic and stellar signatures of deconfinement. We connect the microscopic DM energy-pressure asymmetry to the trace anomaly and use it to discriminate continuously induced DM modifications from the discontinuous thermodynamic response of deconfinement. 

In the pressureless, comoving heavy-WIMP regime, we further identify a universal dark-sector scaling governed by the mass-loading parameter $\lambda=F_\chi m_\chi/m_N$, and verify that distinct $ (F_\chi, m_\chi) $ combinations numerically at fixed $\lambda$ produce overlapping trace-anomaly, sound-speed, and mass--radius responses for a given BM EOS. This establishes that the leading DM-induced stellar response depends on the dark-sector parameters primarily through $\lambda$, while retaining the model dependence of the underlying BM EOS, and connects the microscopic DM energy--pressure asymmetry to a unified diagnostic of smooth DM-induced modifications and genuine first-order deconfinement.

The paper is organized as follows. In \cref{th}, we introduce the theoretical framework, including the hadronic, DM, and quark-matter models, as well as the hybrid-star construction. In \cref{tras}, we provide a general discussion of the trace-anomaly decomposition. In \cref{tras1}, we discuss the conformal dynamics, while in \cref{tras2} we introduce the exact trace-anomaly decomposition. A quantitative assessment of the dark-sector contributions is presented in \cref{tras3}. In \cref{universalR}, we introduce a universal scaling relation in the pressureless, comoving heavy-WIMP limit. We present our results and analysis in \cref{rna}, followed by the main conclusions in \cref{cnc}.

\section{Theoretical framework}\label{th}
The study combines a nonlinear relativistic mean-field model for BM with a WIMP DM model based on the scalar Higgs portal. We further introduce two complementary DM--BM interaction scenarios: a massive vector mediator that couples the two sectors through their densities and a contact current--current interaction. This allows us to examine how different forms of DM--BM coupling affect the dense-matter EOS and to quantify their effects while clearly isolating the model dependence of the results. 
\begin{widetext}
The BM sector is described by:
\begin{align} \label{lgbm}
\mathcal{L}_B &=  \bar{\psi}_B \left[i\gamma^\mu\partial_\mu-\gamma^0 \left(  g_\omega \omega_0 + g_\rho I_{3\rho} \rho_{03} \right)- \left(m_N - g_\sigma \sigma_0 \right)\right] \psi_B - \frac{1}{2} m_\sigma^2 \sigma^2_0 + \frac{1}{2} m_\omega^2 \omega_0^2 + \frac{1}{2} m_\rho^2 \rho_{03}^2 \nonumber\\&
- \frac{1}{3} b\, m_N \, g_\sigma^3 \sigma^3_0- \frac{1}{4} c\, g_\sigma^4 \sigma^4_0+ \frac{\xi}{4!} g_\omega^4\omega_0^4+ \Lambda_\omega g_\rho^2 g_\omega^2 \rho_{03}^2\omega_0^2, \nonumber\\
\mathcal{L}_{l} &=\bar{\psi}_{l}\left(i\gamma^\mu\partial_\mu -m_{l}\right)\psi_{l}.
\end{align}
Here \(\psi_B\) and \(\psi_l\) denote the nucleon and lepton fields, respectively, \(I_3\) is the nucleon isospin projection, and \(\gamma^\mu\) are the Dirac matrices. The quantities \(\sigma_0,\omega_0,\rho_{03}\) denote the mean fields of the \(\sigma\), \(\omega\), and \(\rho\) mesons, with \(g_\sigma,g_\omega,g_\rho\) and \(m_\sigma,m_\omega,m_\rho\) their corresponding couplings and masses. The parameters \(b,c,\xi,\Lambda_\omega\) characterize the nonlinear meson interactions. In particular, $b$ and $c$ control the incompressibility near saturation, $\xi$ softens the EOS at high density, and $\Lambda_\omega$ regulates the density dependence of the symmetry energy \cite{Providencia:2012rx, Tolos:2017lgv}. 

The dark sector is described by
\begin{align}
\mathcal{L}_{\rm HP} =\bar{\chi}\left(i\gamma^\mu\partial_\mu-\left(m_\chi-g_h h\right)\right)\chi+\frac{1}{2}\left(\partial^\mu h\,\partial_\mu h-m_h^2h^2\right)
+\sum_N\frac{f\,m_N}{v}\bar{\psi}_Nh\psi_N.
\label{HP}
\end{align}
\end{widetext}
\Cref{HP} consists of a WIMP fermion~\cite{Arcadi:2017kky} $\chi$ coupled to the Higgs field $h$, with $m_h=125\,\mathrm{GeV}$, $v=246\,\mathrm{GeV}$, and an effective Higgs--nucleon form factor $f=0.35$. We adopt $m_\chi=200\,\mathrm{GeV}$ and $g_h=0.07$, yielding a spin-independent DM--nucleon cross section of order $\sigma_{\chi N}^{\rm SI, H}\simeq 3.9\times10^{-45}\,\mathrm{cm}^2$. The chosen form factor is consistent with lattice-QCD, MILC, and ATLAS determinations \cite{Toussaint:2009pz, ATLAS:2015ciy}, while the coupling lies within the phenomenologically viable range $0.001\leq g_h\leq0.1$ \cite{Panotopoulos:2017idn}.

To obtain a self-consistent DM density, we extend the Higgs-portal framework by incorporating a massive $Z'$ vector mediator ($m_X = 1.5\,\mathrm{TeV}$) associated with a hidden $U(1)_X$ symmetry, described by
\begin{align}
\mathcal{L}_{X}=&-\frac{1}{4} X_{\mu\nu}X^{\mu\nu}+\frac{1}{2} m_X^2X_\mu X^\mu\nonumber\\&-\sum_N g_{NX}\bar{\psi}_N\gamma^\mu\psi_NX_\mu-g_{\chi X}\bar{\chi}\gamma^\mu\chi X_\mu,
\label{eq:Vector}
\end{align}
with $X_{\mu\nu}=\partial_\mu X_\nu -\partial_\nu X_\mu$. In the mean-field limit ($X_\mu \rightarrow X_0$), the mediator generates a vector potential analogous to the RMF $\omega$ field, which self-consistently shifts the chemical potentials and couples BM and DM densities. Such TeV-scale $Z'$ mediators are motivated by theoretical and collider studies~\cite{Langacker:2008yv, CMS:2021dzg, CMS:2020ulv, ATLAS:2024kpy}, with related vector frameworks explored for lighter mediators~\cite{Hardy:2022ufh, Capanelli:2024pzd}. Following ATLAS spin-1 benchmark constraints at $95\%$ CL~\cite{ATLAS:2024kpy, ATLAS:2021ovy, ATLAS:2020fmm}, we set $g_{qX}=0.25$ ($g_{NX}\approx 3g_{qX}=0.75$), $g_{\chi X}=1$, and $m_X=1.5\,\mathrm{TeV}$~\cite{Langacker:2008yv}, yielding a spin-independent DM--nucleon cross section of $\sigma_{\chi N}^{\mathrm{SI, X}}\simeq 1.2\times10^{-41}\,\mathrm{cm}^2$. A detailed derivation of the EOS, including how the dark and visible sectors couple through their respective chemical potentials, can be found in \cite{Issifu:2026ehm}.

It has been shown in Sec.~II.1 of \cite{Issifu:2026diw} that, in the low-energy limit $q\ll m_X$, the $Z'$-mediator interaction in \cref{eq:Vector} reduces to the effective contact interaction
\begin{equation}\label{contact}
\mathcal{L}_{\rm int}=-\alpha(\bar{\chi}\gamma^\mu\chi)(\bar{\psi}_B\gamma_\mu\psi_B),
\end{equation}
with
\begin{equation}
\alpha=\frac{g_{NX}g_{\chi X}}{m_X^2}>0.
\end{equation}
The interaction shifts the effective chemical potentials of the BM and DM sectors as $\mu_{n,p}^{*}=\mu_{n,p}-g_\omega\omega_0-g_\rho I_{3\rho}\rho_{03}-\alpha n_\chi$ and $\mu_\chi^{*}=\mu_\chi-\alpha n_B$. Here, $\mu_{n,p}^{*}=\sqrt{k_{F_{n,p}}^2+m_N^{*2}}$ and $\mu_\chi^{*}=\sqrt{k_{F\chi}^2+m_\chi^{*2}}$ denote the kinetic effective chemical potentials of the BM and DM sectors, respectively. The density-dependent terms explicitly couple the two sectors and determine their coexistence self-consistently. The positive $\alpha$ corresponds to a repulsive vector interaction, which increases the energetic cost of placing DM and BM in the same volume and thereby regulates their coexistence inside the NS. 

The total energy density and pressure are
\begin{align}\label{totPE}
\varepsilon_{\rm tot}&=\varepsilon_B+\varepsilon_\chi+\varepsilon_h+ \varepsilon_X+\varepsilon_{\rm int} \nonumber\\
P_{\rm tot} &=P_B+P_\chi+P_h+ P_X+ P_{\rm int},
\end{align}
with
\begin{equation}
\varepsilon_{\rm int}=\alpha n_B n_\chi, \qquad P_{\rm int}=\alpha n_B n_\chi.
\end{equation}
In the absence of contact interaction, $\varepsilon_{\rm int}=P_{\rm int}=0$. The contact interaction integrates out $X_\mu$, yielding $\varepsilon_X=P_X=0$. Hence, the two contributions are alternative and are shown together in \cref{totPE} solely for compactness. The subscripts $B$, $\chi$, $X$, and $h$ denote BM, DM, vector mediator, and Higgs field, respectively (the full expressions of $\varepsilon_{\rm tot}$ and $P_{\rm tot}$ are presented in Appendix~\ref{apendA}). 

The DM content is characterized by the global particle fraction $F_\chi=N_\chi/N_B$, introduced for mirror DM in \cite{Issifu:2026diw}. Unlike approaches that infer the global fraction by integrating separate density profiles \cite{Hajkarim:2024ecp,1996csnp.bookGlen}, we prescribe $F_\chi$ and impose the local closure $n_\chi(r)=F_\chi n_B(r)$, so $n_\chi$ and Fermi momentum are determined self-consistently from $n_B$. This prescription is motivated by efficient capture, thermalization, and spatial mixing, making the DM approximately comoving with the BM.

\subsection{Hybrid NS admixed with DM}
To study the behavior of the trace anomaly across deconfinement, we construct hybrid NS by matching the hadronic EOS to an independent quark-matter (QM) EOS through a Maxwell construction. The transition is determined by chemical and mechanical equilibrium,
\begin{equation}
    P_H(\mu_B)=P_Q(\mu_B)\equiv P_{\rm cr}, \qquad \mu_B^H=\mu_B^Q\equiv\mu_{\rm cr},
\end{equation}
with the associated discontinuity in the energy density characterizing the first-order transition \cite{Glendenning:1992vb}. For the quark phase, we employ the MDDQM model introduced in \cite{Issifu:2025byg}, in which
\begin{equation}
    m_q(n_B)  =m_{q0}+\frac{D}{n_B^{1/3}} +\left(1+\kappa\, n_B^{1/3}\right)n_B^{1/3},
\end{equation}
where $m_{q0}$ is the bare mass and $q=u,d,s$ labels the quark flavors. The parameters $D$ and $\kappa$ were optimized in \cite{Issifu:2025byg} by imposing the Bodmer--Witten conjecture together with astrophysical constraints. We consider two representative parameterizations,
\begin{equation*}
\begin{aligned}
{\rm MDDQM1}:&\quad
\kappa=0.0031~\mathrm{MeV}^{-1},
\quad \sqrt{D}=108.54~\mathrm{MeV},\\
{\rm MDDQM2}:&\quad
\kappa=0.0045~\mathrm{MeV}^{-1},
\quad \sqrt{D}=102.85~\mathrm{MeV},
\end{aligned}
\end{equation*}
which yield maximum quark-star (QS) masses of $2.30\,M_\odot$ and $2.37\,M_\odot$, respectively, and therefore satisfy the $\sim2\,M_\odot$ massive-pulsar constraint. The details for how the QM is coupled to the dark sector have been detailed in Appendix~\ref{apendB}.

\section{Trace anomaly decomposition}\label{tras}
\begin{table}[t!]
\small
\centering
\caption{Structural properties of DM admixed NSs predicted by the NL3$\omega\rho$, DD2, and FSU2R RMF models for different $F_\chi$. The quantities $\varepsilon_c$, $P_c$, and $n_c$ denote the central energy density, pressure, and baryon density of the maximum-mass configuration, respectively. The quantities $w_\chi$ and $\delta\Delta_c$ are the estimated central DM energy-density fraction and the corresponding shift in the trace anomaly, respectively, evaluated using a heavy-WIMP approximation with $m_\chi=200~\mathrm{GeV}$.}
\begin{ruledtabular}
\begin{tabular}{ccccccc}
Model & $F_\chi$ & $\varepsilon_c$& $P_c$  & $n_c$ & $w_\chi$ & $\delta\Delta_c$ \\
 & (\%) &  (MeV/fm$^{3}$)&  (MeV/fm$^{3}$) & (fm$^{-3}$)&  & \\
\hline

\multicolumn{7}{c}{\textbf{NL3$\omega\rho$}}\\
\hline
NL3$\omega\rho$ & 0.00 & 889.95 & 460.85 & 0.69 & 0 & 0   \\
& 0.05 & 1022.16 & 507.81 & 0.72 & 0.07 & 0.04  \\
& 0.10 & 1160.29 & 556.70 & 0.75 & 0.13 & 0.06 \\
& 0.20 & 1466.14 & 669.21 & 0.81 & 0.22 & 0.10 \\
\hline

\multicolumn{7}{c}{\textbf{DD2}}\\
\hline
DD2 & 0.00 & 1067.54 & 499.02 & 0.83 & 0 & 0  \\
& 0.05 & 1239.22 & 562.12 & 0.88 & 0.07 & 0.03  \\
& 0.10 & 1424.71 & 628.58 & 0.92 & 0.13 & 0.06 \\
& 0.20 & 1815.41 & 766.52 & 1.01 & 0.22 & 0.09\\
\hline

\multicolumn{7}{c}{\textbf{FSU2R}}\\
\hline
FSU2R & 0.00 & 1152.39 & 343.01 & 0.94 & 0 & 0\\
& 0.05 & 1343.82 & 380.27 & 1.00 & 0.07 & 0.02 \\
& 0.10 & 1539.15 & 412.19 & 1.05 & 0.14 & 0.04 \\
& 0.20 & 1967.35 & 480.98 & 1.16 & 0.24 & 0.06 \\
\end{tabular}
\end{ruledtabular}
\label{tab:maxmass}
\end{table}

\begin{table*}[t!]
\centering
\caption{Critical pressure $P_{\rm cr}$ and chemical potential $\mu_{\rm cr}$ at the hadron-quark phase transition for the NL3$\omega\rho$ model at different $F_\chi$. $n_c$, $\varepsilon_c$ and $P_c$ denote the baryon density, energy density and pressure at the center of the most massive stable star with mass $M_{\rm max}$ and radius $R$. The quantities $\Delta_{\rm tot}$ and $\delta\Delta_c$ denote the total trace anomaly at the corresponding central density and its shift relative to the $F_\chi=0$ configuration of the same MDDQM parametrization, respectively.}
\begin{ruledtabular}
\begin{tabular}{ccccccccccc}
Model & $F_\chi$ & $P_{\rm cr}$ & $\mu_{\rm cr}$
& $n_c$ & $\varepsilon_c$ 
& $P_c$ & $\Delta_{\rm tot}$
& $\delta\Delta_c$ & $M_{\max}$ & $R$ \\
 & (\%) & (MeV fm$^{-3}$) &  (MeV)
& (fm$^{-3}$) &  (MeV fm$^{-3}$)
&(MeV fm$^{-3}$) &
& & $(M_\odot)$ & (km) \\
\hline
MDDQM1 
& 0.00 & 177.43 & 1474.94 & 0.58 & 761.34 & 218.69 & 0.0461 & 0.00 & 2.58 & 13.88 \\
& 0.05 & 177.42 & 1474.93 & 0.59 & 853.12 & 224.57 & 0.0701 & 0.02 & 2.35 & 12.98 \\
& 0.10 & 177.42 & 1474.94 & 0.62 & 971.30 & 243.76 & 0.0824 & 0.04 & 2.14 & 12.16 \\
& 0.20 & 177.43 & 1474.94 & 0.68 & 1209.40 & 279.54 & 0.1022 & 0.06 & 1.82 & 10.86 \\
\hline
MDDQM2 
& 0.00 & 201.47 & 1524.46 & 0.58 & 787.66 & 253.35 & 0.0117 & 0.00 & 2.63 & 13.76 \\
& 0.05 & 201.46 & 1524.44 & 0.59 & 889.29 & 263.62 & 0.0369 & 0.03 & 2.40 & 12.86 \\
& 0.10 & 201.46 & 1524.44 & 0.61 & 992.66 & 281.22 & 0.0500 & 0.04 & 2.20 & 12.10 \\
& 0.20 & 201.46 & 1524.44 & 0.67 & 1243.94 & 323.50 & 0.0733 & 0.06 & 1.89 & 10.80 \\
\end{tabular}
\end{ruledtabular}
\label{tab:critical_values}
\end{table*}

\subsection{Conformal dynamics}\label{tras1}
The trace anomaly quantifies the breaking of conformal symmetry in dense matter. Following \cite{Fujimoto:2022ohj,Annala:2023cwx}, we use the normalized trace anomaly as
\begin{equation}\label{traceAn}
\Delta_{\rm tot}\equiv\frac{\varepsilon_{\rm tot}-3P_{\rm tot}}{3\varepsilon_{\rm tot}}=\frac{I_{\rm tot}}{3\varepsilon_{\rm tot}}.
\end{equation}
In the following, we introduce $I_i=\varepsilon_i-3P_i$ to denote the interaction measure of each sector $i$, indexing baryons and DM species. Using \cref{engd,pres}, the individual contributions are
\begin{equation}
I_X=-m_X^2X_0^2,\quad I_{\rm int}=-2\alpha F_\chi n_B^2,\quad I_h=2m_h^2h_0^2,
\end{equation}
together with $I_B=\varepsilon_B-3P_B$ and $I_\chi=\varepsilon_\chi-3P_\chi$. Hence,
\begin{equation}
I_{\rm tot}=I_B+I_\chi+2m_h^2h_0^2-m_X^2X_0^2-2\alpha F_\chi n_B^2.
\label{eq:I_final}
\end{equation}
The massive-vector and contact-interaction terms represent alternative DM--BM couplings. For the vector model, $X_0\neq0$ and $-m_X^2X_0^2$ is retained, while the contact term is omitted. In the contact limit, $X_\mu$ is integrated out, so $X_0=0$, and $-2\alpha F_\chi n_B^2$ encodes its effect. Thus, the two contributions are treated separately, not simultaneously. The density derivative of the trace anomaly is then
\begin{equation}
\frac{d\Delta_{\rm tot}}{dn_B}=\frac{I'_{\rm tot}\varepsilon_{\rm tot}-I_{\rm tot}\varepsilon'_{\rm tot}}{3\varepsilon_{\rm tot}^2},
\label{eq:dDelta}
\end{equation}
where primes denote derivatives with respect to $n_B$.

These expressions separate the microscopic contributions to conformal-symmetry breaking: the BM and DM sectors enter through their kinetic and mean-field dynamics, while the Higgs, vector, and contact interactions contribute with positive, negative, and negative interaction measures, respectively. Thus, $\Delta_{\rm tot}$ and $d\Delta_{\rm tot}/dn_B$ provide complementary diagnostics of the DM-induced deviation of conformality.






\subsection{Exact trace-anomaly decomposition}\label{tras2}
Within the present self-consistent framework, substituting \cref{totPE} into \cref{traceAn} yields the exact identity
\begin{equation}
\Delta_{\rm tot}(n_B)=\sum_a w_a(n_B)\,\Delta_a(n_B),
\label{eq:Delta_identity}
\end{equation}
where
\begin{equation}
w_a=\frac{\varepsilon_a}{\varepsilon_{\rm tot}}, \quad \Delta_a=\frac{1}{3}-\frac{P_a}{\varepsilon_a}, \quad\sum_a w_a=1.
\end{equation}
Thus, $\Delta_{\rm tot}$ is the energy-density-weighted average of the trace anomalies of the individual sectors. The contribution of each sector is therefore determined by both its intrinsic conformal properties, $\Delta_a$, and its fractional contribution to the total energy density, $w_a$.

The BM sector provides the reference scale for $\Delta_B$. At low densities, where $P_B\ll\varepsilon_B$, one has $\Delta_B\simeq1/3$. In the conformal limit, $P_B=\varepsilon_B/3$, so $\Delta_B\to0$ signals the approach to conformality. At sufficiently high densities, the repulsive vector field can become dominant, with $P_\omega\simeq\varepsilon_\omega$, giving $\Delta_\omega\simeq-2/3$. 

\subsection{Dark-sector contribution}\label{tras3}
Using the DM mass benchmark $m_\chi=200~\mathrm{GeV}$, the largest central baryon density in \cref{tab:maxmass} is $n_c=1.16~\mathrm{fm}^{-3}$ for the FSU2R model with $F_\chi=0.2\%$. Estimating the DM Fermi momentum from $k_F^\chi=(3\pi^2F_\chi n_c)^{1/3}$, we obtain $k_F^\chi\simeq80.81~\mathrm{MeV}$. Thus, the DM remains extremely nonrelativistic even at the center of the most compact configuration considered, with ${k_F^\chi}/{m_\chi^*}\simeq4.04\times10^{-4}$, where $m_\chi^*\simeq m_\chi$ has been used. Expanding the Fermi integrals in \cref{engd,pres} in powers of $k_F^\chi/m_\chi^*$ gives
\begin{align}
\varepsilon_\chi&\simeq n_\chi m_\chi^* +\frac{3}{5}n_\chi\frac{(k_F^\chi)^2}{2m_\chi^*}+\cdots, \label{energ_L}
\\
P_\chi&\simeq\frac{2}{5}n_\chi\frac{(k_F^\chi)^2}{2m_\chi^*}+\cdots, \label{press_L}
\end{align}
where
\begin{equation}
n_\chi=\frac{\gamma_\chi}{2\pi^2}\int_0^{k_F^\chi}k^2\,dk = \dfrac{\gamma_\chi (k_F^\chi)^3}{6\pi^2}.
\end{equation}
Since $P_\chi\ll\varepsilon_\chi$, the DM contribution satisfies $\Delta_\chi\simeq1/3$. Unlike $\Delta_B$, which evolves significantly with $n_B$, the heavy-WIMP contribution therefore remains close to its nonrelativistic limit throughout the star.

The Higgs and vector fields follow directly from \cref{engd,pres}, giving $P_h\simeq-\varepsilon_h$ and $P_X\simeq\varepsilon_X$, and hence $\Delta_h\simeq4/3$ and $\Delta_X\simeq-2/3$, respectively. From \cref{engd}, $\varepsilon_X=(1/2)m_X^2X_0^2$; substituting \cref{veom} yields
\begin{align}\label{vself}
\varepsilon_X
&= \frac{g_{NX}^2 n_B^2}{2m_X^2}+ \frac{g_{\chi X}^2 n_\chi^2}{2m_X^2}+ \frac{g_{NX}g_{\chi X}n_B n_\chi}{m_X^2}.
\end{align}
Similarly, $\varepsilon_h=(1/2)m_h^2h^2$. Substituting the mean-field equation of motion for $h$ gives
\begin{align}\label{scself}
\varepsilon_h
&= \frac{f^2m_N^2}{2v^2m_h^2}(n_s^B)^2+ \frac{g_h^2(n_s^\chi)^2}{2m_h^2}+ \frac{fm_Ng_h}{vm_h^2}n_s^B n_s^\chi.
\end{align}
The interaction in \cref{vself} depends explicitly on the local densities, while the EOS determines their stellar profiles. For the benchmark $g_{NX}g_{\chi X}=0.75$ and $n_\chi=F_\chi n_B$, we obtain at the center of the most massive stable star $\varepsilon_{X} \simeq 1.30\times10^{-6}~\mathrm{MeV\,fm^{-3}}$.

For the Higgs contribution, the heavy-WIMP limit allows the DM scalar density to be expanded as
\begin{align}
n_s^\chi &=\frac{\gamma_\chi}{2\pi^2}\int_0^{k_F^\chi}\frac{m_\chi^*}{\sqrt{k^2+m_\chi^{*2}}}\,k^2\,dk
\nonumber\\
&\simeq\frac{\gamma_\chi}{2\pi^2}\left[\frac{(k_F^\chi)^3}{3}-\frac{(k_F^\chi)^5}{10m_\chi^{*2}}+\frac{3(k_F^\chi)^7}{56m_\chi^{*4}}+\cdots\right].
\end{align}
The leading term is $n_\chi$, while $n_s^B\simeq n_B$ at leading order, giving $\varepsilon_h\simeq 7.21\times10^{-10}~\mathrm{MeV\,fm^{-3}}$. Since $n_s^B/n_B<1$ in the NS core, this is an upper-limit estimate. With $\varepsilon_c\sim10^{3}~\mathrm{MeV\,fm^{-3}}$, one finds $w_h\ll10^{-10}$, $w_X\sim 10^{-9}$, while the contact interaction, $\varepsilon_{\rm int}\simeq6.88\times 10^{-9}~\mathrm{MeV\,fm^{-3}}$, leads to $w_{\rm int}\sim10^{-11}$. Thus, the Higgs, vector, and contact contributions to $\Delta_{\rm tot}$ are negligible for the adopted benchmarks.

Neglecting the numerically negligible Higgs, vector, and contact contributions, the exact decomposition in \cref{eq:Delta_identity} reduces to $\Delta_{\rm tot}=w_\chi\Delta_\chi+w_B\Delta_B$ with $w_\chi+w_B=1$. In the heavy-WIMP limit, $\Delta_\chi\simeq1/3$, giving
\begin{equation}
\Delta_{\rm tot}(n_B)
\simeq
(1-w_\chi)\Delta_B(n_B)+\frac{w_\chi}{3}.
\label{eq:DeltaReduced}
\end{equation}
Using the leading nonrelativistic contribution \cref{energ_L} $\varepsilon_\chi\simeq F_\chi n_Bm_\chi^*$, the DM energy-density fraction is
\begin{equation}
w_\chi=\frac{\varepsilon_\chi}{\varepsilon_{\rm tot}}
\simeq
\frac{F_\chi n_Bm_\chi^*}{\varepsilon_{\rm tot}},
\end{equation}
where $\varepsilon_\chi = n_\chi m_\chi$ is precisely the rest mass energy of the WIMP particle. The shift relative to $\Delta_B$ is then
\begin{equation}
\Delta_{\rm tot}-\Delta_B = w_\chi\left(\frac13-\Delta_B\right)= w_\chi\frac{P_B}{\varepsilon_B},
\label{eq:Shift}
\end{equation}
where the last equality follows directly from $\Delta_B=1/3-P_B/\varepsilon_B$. In the high-$n_B$ regime where $\Delta_B\leq0$, DM shifts $\Delta_{\rm tot}$ away from $\Delta_B$, with the effect reaching $\Delta_{\rm tot}-\Delta_B\sim0.1$ over the considered $F_\chi$ range. The corresponding central shifts and $w_\chi$ for the different EOSs are given in \cref{tab:maxmass}.

\section{Universal DM scaling in the pressureless, comoving heavy-WIMP limit} \label{universalR}
From \cref{energ_L,press_L} we have established that, for sufficiently heavy DM, the dark particles are non-relativistic and their pressure is much smaller than their rest-mass energy density, i.e.,
\begin{equation}
P_\chi \ll \varepsilon_\chi,
\qquad
\varepsilon_\chi \simeq m_\chi n_\chi.
\label{eq:NR_DM}
\end{equation}
In this limit, we consider DM to be in a pressureless regime. We further demonstrated that the effective DM--BM interactions are negligible compared with the BM and DM contributions. Following these assumptions, and the comoving character of DM, i.e., $n_\chi=F_\chi n_B$, we get
\begin{equation}\label{scale}
    \varepsilon_{\rm tot} \simeq\varepsilon_B + F_\chi n_B m_\chi,\quad P_{\rm tot}\simeq P_B.
\end{equation} 
In this limit, $F_\chi$ and $m_\chi$ enter the EOS only through their product $F_\chi m_\chi$ motivating the introduction of the mass-loading parameter
\begin{equation}
\lambda \equiv F_\chi\frac{m_\chi}{m_N}.
\label{eq:lambda}
\end{equation}
Therefore, we rewrite the $\varepsilon_{\rm tot}$ and $P_{\rm tot}$ in terms of $\lambda$ as 
\begin{align}
\varepsilon_{\rm tot}(n_B) &\simeq \varepsilon_B(n_B)+\lambda m_N n_B, \label{denapprox}\\ 
P_{\rm tot}(n_B)&\simeq P_B(n_B).
\label{eq:lambda_EOS}
\end{align}
Thus, DM adds an energy per baryon equal to a fraction $\lambda$ of the nucleon rest-mass energy. This defines a one-parameter scaling class for pressureless, comoving DM. The universality concerns the dependence on the dark-sector parameters: within this limit, the static stellar structure depends on $F_\chi$ and $m_\chi$ only through the mass-loading parameter $\lambda$, while the underlying BM EOS remains model-dependent.

\subsection{Thermodynamic response of the stellar matter}
Using the relation $d\varepsilon_B/dn_B=\mu_B$, where $\mu_B$ is the chemical potential of the BM, we obtain 
\begin{equation}\label{tchem}
\frac{d\varepsilon_{\rm tot}}{dn_B} =\mu_B+\lambda m_N.
\end{equation}
Since 
\begin{equation}\label{pderiv}
\frac{dP_B}{dn_B}=\frac{dP_B}{d\varepsilon_B}\frac{d\varepsilon_B}{dn_B}=c_{s,B}^2\mu_B,
\end{equation}
we can write
\begin{equation}
c_{s,\rm tot}^2=\frac{dP_{\rm tot}/dn_B}{d\varepsilon_{\rm tot}/dn_B}= c_{s,B}^2\frac{\mu_B}{\mu_B+\lambda m_N},
\end{equation}
equivalently, 
\begin{equation}
\frac{c_{s,\rm tot}^2}{c_{s,B}^2} = \frac{1} {1+\lambda m_N/\mu_B}.
\label{eq:cs_ratio}
\end{equation}
Since in the presence of DM $\lambda>0$, this expression implies that $c_{s,\rm tot}^2<c_{s,B}^2$; consequently, the suppression of sound speed does not require a significant microscopic DM--BM interaction. It is controlled by the ratio of the DM energy loading per baryon, $\lambda m_N$, to the baryonic chemical potential $\mu_B$. For a weak DM loading $\lambda m_N/\mu_B\ll1$, we expand \cref{eq:cs_ratio} as 
\begin{equation}
c_{s,\rm tot}^2\simeq c_{s,B}^2 \left( 1-\frac{\lambda m_N}{\mu_B}\right) +\mathcal{O}(\lambda^2).
\end{equation}
Thus, in the pressureless-comoving regime {($P_\chi\simeq 0$, $n_\chi=F_\chi n_B$)}, the leading modification of the sound speed is controlled by the single DM mass-loading parameter $\lambda$.

The adiabatic index, $\Gamma$, is also expressed as 
\begin{equation}\label{adiab}
    \Gamma_{\rm tot}= \dfrac{P_{\rm tot}+ \varepsilon_{\rm tot}}{P_{\rm tot}}\dfrac{dP_{\rm tot}}{d\varepsilon_{\rm tot}}=\dfrac{P_{\rm tot}+ \varepsilon_{\rm tot}}{P_{\rm tot}}c_{s, tot}^2,
\end{equation}
where $\Gamma_{\rm tot}$ is the total adiabatic index \cite{Paschalidis:2016vmz}. Using the pressureless and comoving assumption, we can rewrite \cref{adiab} using \cref{scale,tchem,pderiv} as
\begin{align}
    \Gamma_{\rm tot}= \dfrac{P_B+ \varepsilon_B+\lambda m_Nn_B}{P_B}\dfrac{c^2_{s,B}\mu_B}{\mu_B + \lambda m_N},
\end{align}
using the definition of enthalpy density, $P_B + \varepsilon_B = n_B\mu_B$. This further reduces the expression to 
\begin{align}
    \Gamma_{\rm tot} &= \dfrac{n_B(\mu_B+\lambda m_N)}{P_B}\dfrac{c^2_{s,B}\mu_B}{\mu_B+\lambda m_N}\nonumber\\ &=\dfrac{P_B+\varepsilon_B}{P_B}c_{s,B}^2= \Gamma_B,
\end{align}
where $\Gamma_B$ is the adiabatic index of the BM without DM. Also, this equivalence becomes clear by defining $\Gamma_{\rm tot}$ in terms of $P_{\rm tot}$ and $n_B$ as $\Gamma_{\rm tot}=(n_B/P_{\rm tot})(dP_{\rm tot}/dn_B)$, since $P_{\rm tot}=P_B$ in the pressureless-comoving limit.

Although the $c^2_{s,\rm tot}$ is suppressed according to \cref{eq:cs_ratio}, the $\Gamma_{\rm tot}$ remains exactly unchanged in the pressureless-comoving limit. Indeed, the DM-induced increase in $\varepsilon_{\rm tot}+P_{\rm tot}=n_B(\mu_B+\lambda m_N)$, is exactly compensated by the corresponding reduction of $c_{s,\rm tot}^2$, yielding $\Gamma_{\rm tot}=\Gamma_B$. Thus, the DM admixture does not modify the intrinsic logarithmic pressure response of the stellar matter, since $P_{\rm tot}=P_B$, but it does reduce the propagation speed of pressure perturbations by increasing the energy density that must be accelerated. Physically, $\lambda$ changes the dynamical propagation of perturbations without altering the underlying adiabatic stiffness of the BM EOS.

\subsection{Trace-anomaly scaling and universality}
Using \cref{denapprox} we can express \cref{eq:Shift} as
\begin{equation}
\Delta_{\rm tot}-\Delta_B =\frac{\lambda m_N n_B}{\varepsilon_B+\lambda m_N n_B}\frac{P_B}{\varepsilon_B}.
\label{eq:Delta_lambda}
\end{equation}
A useful diagnostic for testing universality is
\begin{equation}
\mathcal{U}_\chi(n_B) \equiv \frac{\Delta_{\rm tot}-\Delta_B} {w_\chi\left(\frac{1}{3}-\Delta_B\right)} \equiv \frac{\Delta_{\rm tot}-\Delta_B} {w_\chi (P_B/\varepsilon_B)} .
\label{eq:universality_test}
\end{equation}
The pressureless-comoving limit predicts $\mathcal{U}_\chi=1$, see \cref{eq:Shift}). Thus, $\mathcal{U}_\chi$ provides a direct test of the pressureless-comoving scaling in the full numerical solutions. Deviations from $\mathcal{U}_\chi=1$ quantify corrections associated with DM pressure, relativistic effects, or non-negligible interaction contributions.

At leading order in the DM loading, and approximating $\varepsilon_B\simeq m_N n_B$, one obtains
\begin{equation}
w_\chi\simeq F_\chi\frac{m_\chi}{m_N}=\lambda,
\label{eq:wx_approx}
\end{equation}
for small $\lambda$. In such a leading-order expansion, \cref{eq:Shift} becomes
\begin{equation}
    \lambda = \dfrac{\Delta_{\rm tot}-\Delta_B}{{1}/{3}-\Delta_B}.
\end{equation}
Consequently, we can define a DM-fraction diagnostic parameter
\begin{equation}
\mathcal{I}_\chi \equiv \frac{m_N}{m_\chi}\frac{\Delta_{\rm tot}-\Delta_B}{\frac{1}{3}-\Delta_B},
\label{eq:dark_invariant}
\end{equation}
one obtains $\mathcal{I}_\chi\simeq F_\chi$. Thus, in the heavy, nonrelativistic regime, the leading conformal response is governed by the single DM mass-loading parameter $\lambda=F_\chi m_\chi/m_N$, while microscopic interaction details enter only through subleading corrections to the EOS and self-consistent field dynamics.

\section{Results and analysis}\label{rna}

\Cref{tab:maxmass} shows that, in the purely hadronic DM admixed sector, increasing $F_\chi$ from $0$ to $0.2\%$ raises $w_\chi$ from 0 to $\simeq0.22$--$0.24$ and produces a systematic upward shift in the trace anomaly, reaching $\delta\Delta_c\simeq0.10$, $0.09$, and $0.06$ for NL3$\omega\rho$, DD2, and FSU2R, respectively. The ordering $\mathrm{NL3}\omega\rho>\mathrm{DD2}>\mathrm{FSU2R}$ reflects the different high-density behavior of the underlying BM EOSs, showing that the magnitude of the DM-induced conformality shift depends not only on $F_\chi$ but also on the BM response. 

\Cref{tab:critical_values} demonstrates that this DM-induced modification persists in the hybrid NS sector without appreciably changing the transition point: for each MDDQM parametrization, $P_{\rm cr}$ and $\mu_{\rm cr}$ vary only marginally across the full $F_\chi$ range, whereas the $\varepsilon_c$ and $\Delta_{\rm tot}$ increase and $M_{\max}$ decreases from $\sim2.6\,M_\odot$ to $\sim1.8$--$1.9\,M_\odot$. Thus, the two tables separate the effects of the two physical mechanisms: DM produces a smooth, EOS-dependent modification of conformality, while the first-order hadron--quark transition retains essentially fixed coexistence conditions and generates the distinct discontinuous thermodynamic signature identified in the hybrid star analysis.

\subsection{DM-induced deviation from conformality}
\begin{figure}[t!]
   \centering	
	\includegraphics[width=0.48\textwidth]{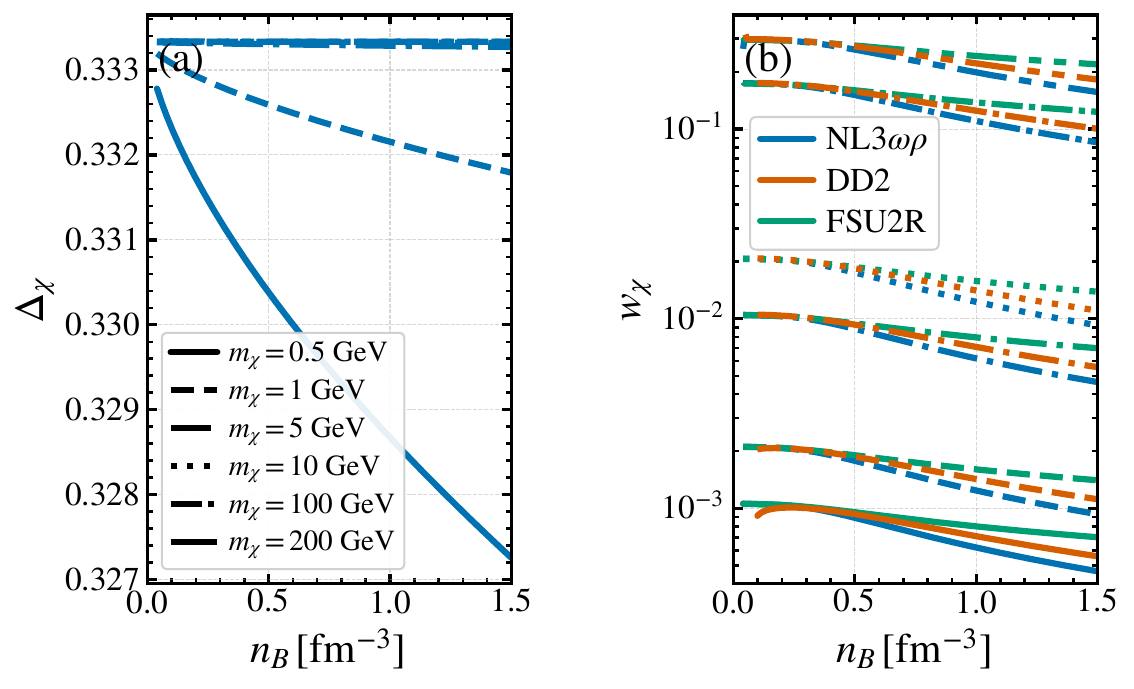}
    	\caption{Density dependence of the DM trace anomaly $\Delta_\chi$ (a) and energy-density weight $w_\chi=\varepsilon_\chi/\varepsilon_{\rm tot}$ (b) for $F_\chi=0.2\%$. In panel~(a), the curves show the dependence on the DM mass $m_\chi$, while in panel~(b) colors distinguish the NL3$\omega\rho$, DD2, and FSU2R EOSs, and line styles denote $m_\chi$.}
\label{fig:mass_weight}
\end{figure}

\Cref{fig:mass_weight} supports the heavy-WIMP approximation used in the benchmark analysis. In panel~(a), $\Delta_\chi(n_B)$ remains close to $1/3$ for $m_\chi\gtrsim5$~GeV, while only the $0.5$ and $1$~GeV cases show a noticeable $n_B$-dependence. Thus, the nonrelativistic limit is reached at $\sim 5$ GeV, well below the $200$~GeV benchmark. 
Panel~(b) shows that $w_\chi$ increases strongly with $m_\chi$, from $\sim10^{-3}$ at $m_\chi=0.5$~GeV to a few $\times10^{-1}$ at $m_\chi=200$~GeV, whereas its $n_B$-dependence is comparatively mild across the three EOSs. The two panels therefore show that $m_\chi$ primarily determines both the intrinsic conformal response and energetic importance of the dark component, with the EOS producing only a modest modulation of $w_\chi$.

\begin{figure*}[t!]
   \centering	
	\includegraphics[width=1\textwidth]{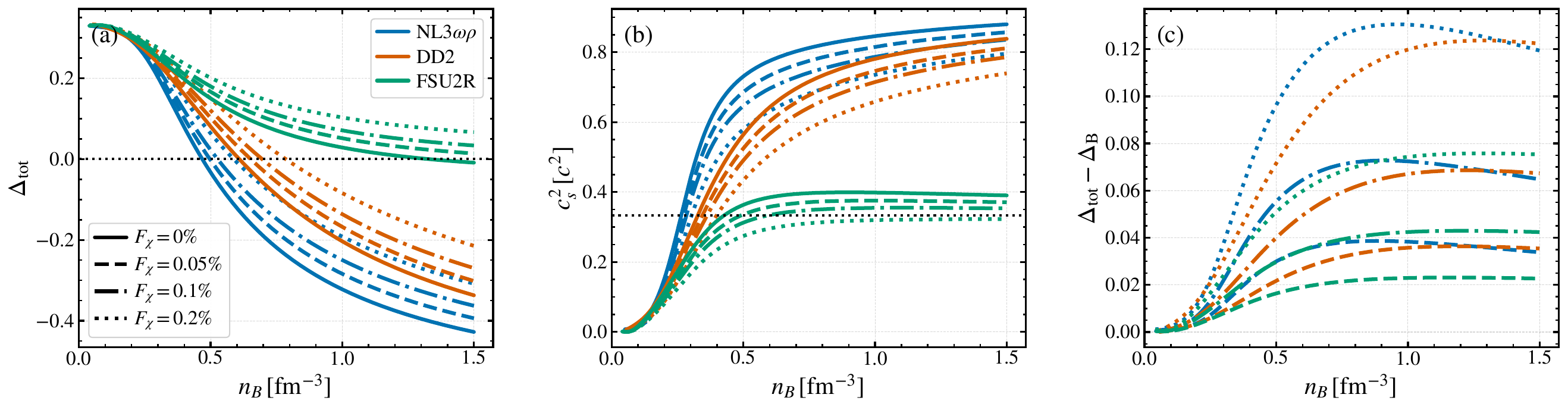}
    	\caption{Conformality diagnostics for DM-admixed NS matter with the NL3$\omega\rho$, DD2, and FSU2R EOSs for different $F_\chi$. Panels~(a), (b), and (c) show $\Delta_{\rm tot}$, squared speed of sound, and the DM-induced shift $\Delta_{\rm tot}-\Delta_{\rm B}$, respectively. Colors denote the EOSs, while line styles denote the DM fractions.}
\label{fig:conformality3panel}
\end{figure*}
\Cref{fig:conformality3panel} summarizes the $n_B$-dependence of $\Delta_{\rm tot}$ and its DM-induced modification for the three RMF models. Panel~(a) shows that all models approach $\Delta_{\rm tot}\simeq 1/3$ at low $n_B$ before diverging at higher $n_B$: for $F_\chi=0\%$, NL3$\omega\rho$ departs most strongly from conformality, reaching $\Delta_{\rm tot}\sim-0.4$, followed by DD2 ($\sim-0.3$), while FSU2R remains closest to $\Delta_{\rm tot}\simeq0$.
These results are in agreement with the expected behavior of the models considered. The model NL3$\omega\rho$ does not contain the $\omega^4$ term in the Lagrangian. As a result, in the large baryonic density limit, the speed of sound approaches 1, as shown in panel (b), and the normalized trace anomaly approaches -2/3. On the other hand, the parametrization FSU2R includes a $\omega^4$ term. In this case, in the large density limit,  the speed of sound squared approaches the conformal limit 1/3, and the trace anomaly approaches 0 (for a discussion of the large density behavior of the speed of sound in RMF models, see the reference \cite{Mueller:1996pm}). The DD2 model has density-dependent couplings, and in the large density limit, the speed of sound also approaches 1. 

Increasing $F_\chi$ systematically shifts the curves in panel (a) upward, reducing the magnitude of the negative $\Delta_{\rm tot}$ for models NL3$\omega\rho$ and DD2, with the effect becoming most pronounced for the EOSs that develop the largest negative $\Delta_{\rm B}$, which implies that for these models the fraction $\omega_\chi$ is larger. Panel~(b) shows the corresponding squared speed of sound. The $c_s^2$ decreases systematically with increasing $F_\chi$, with the separation relative to the no-DM baseline becoming progressively more pronounced toward the high-density region. Similar to panel~(a), increasing $F_\chi$ drives $c_s^2$ toward the conformal limit, $c_s^2=1/3$, indicated by the black dotted line. The suppression of $c_s^2$ follows the same ordering as in panel~(a), NL3$\omega\rho$ $>$ DD2 $>$ FSU2R. 

Panel~(c) isolates the DM-induced contribution through $\Delta_{\rm tot}-\Delta_{\rm B}$, which grows rapidly from low $n_B$, approaches a plateau at $n_B\sim0.75$--$1.0~\mathrm{fm}^{-3}$, and remains nearly constant thereafter, consistent with the competing $n_B$ dependence of $w_\chi$ and $P_{\rm B}/\varepsilon_{\rm B}$ in \cref{eq:Shift}.
NL3$\omega\rho$ exhibits the largest relative shift, reaching $\sim0.12$ for $F_\chi=0.2\%$, because its BM sector develops the most negative $\Delta_B$ and hence allows the largest relative upward shift toward the conformal limit. Thus, the DM-induced shift is smooth with $n_B$ and approaches a saturation value at high $n_B$, with its magnitude controlled by both {$w_\chi$,  or $F_\chi$  according to (\cref{eq:wx_approx}),} and the high-$n_B$ behavior of the underlying BM EOS, {i.e. the large $n_B$ behavior of the speed of sound.}.

\subsection{DM effects on conformality and deconfinement}

\begin{figure*}[t!]
   \centering	
	\includegraphics[width=1\textwidth]{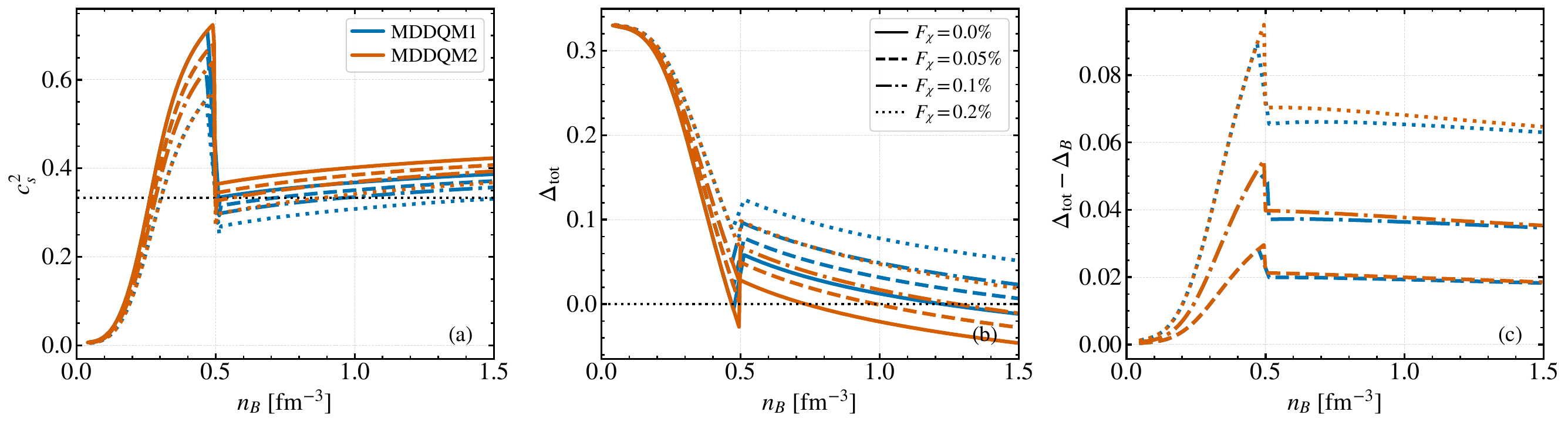}
    	\caption{Conformality diagnostics for hybrid NS obtained with the NL3$\omega\rho$ hadronic model combined with the two modified DDQM parameterizations, MDDQM1 and MDDQM2, for different $F_\chi$. Panel~(a) shows the squared sound speed, $c_s^2$, panel~(b) the total trace anomaly, $\Delta_{\rm tot}$, and panel~(c) the DM-induced shift, $\Delta_{\rm tot}-\Delta_B$, as functions of $n_B$. The colors distinguish MDDQM1 and MDDQM2, while the line styles denote the different values of $F_\chi$.}
\label{hybrid_stars}
\end{figure*}
\Cref{hybrid_stars} shows the conformality diagnostics of hybrid stars obtained with the NL3$\omega\rho$ hadronic model and the MDDQM1 and MDDQM2 quark-matter parametrizations for different $F_\chi$. In panel~(a), $c_s^2=\partial P_{\rm tot}/\partial\varepsilon_{\rm tot}$ increases monotonically along the hadronic branch, as in the pure hadronic EOS, but its magnitude decreases with increasing $F_\chi$ before dropping sharply at $n_B\simeq0.5~\mathrm{fm}^{-3}$, signaling the first-order hadron--quark transition. Panel~(b) exhibits the corresponding upward discontinuity in $\Delta_{\rm tot}$, whereas panel~(c) shows the DM-induced shift, $\Delta_{\rm tot}-\Delta_{\rm B}$, with a visible discontinuity induced by the phase transition mechanism that grows with $F_\chi$. 
For each MDDQM parametrization, $P_{\rm cr}$ and $\mu_{\rm cr}$ remain essentially unchanged with $F_\chi$, indicating that DM contributes negligibly to the pressure balance governing phase coexistence, while its $\varepsilon_\chi$ substantially increases the $\varepsilon_{\rm tot}$ and hence, modifies the thermodynamic response. 

It is important to note that, from \cref{tchem}, the total chemical potential, $\mu_{\rm tot}=\mu_B+\lambda m_N$, acquires a constant shift in the pressureless, heavy-DM limit. The quark phase receives the same shift because its total energy density is given by $\varepsilon_{\rm tot}^{Q}=\varepsilon_Q+\lambda m_N n_Q$, where $n_Q$ denotes the baryon density in the quark phase, as defined in \cref{qbaden}.
Therefore, the Maxwell coexistence conditions, $P_H=P_Q$, and $\mu_B^H=\mu_B^Q$, remain unchanged when $P_\chi\simeq0$ and $\mu_\chi=\lambda m_N\simeq\mathrm{const.}$. Consequently, $P_{\rm cr}$ and $\mu_{\rm cr}$ remain unchanged, since $\lambda m_N$ contributions cancels when the two phases are matched: $\mu_B^{H}+\lambda m_N=\mu_B^{Q}+\lambda m_N$ which implies $\mu_B^{H}=\mu_B^{Q}=\mu_{\rm cr}$.  In contrast, the energy density at the transition is modified by $\lambda m_N$. The energy-density discontinuity therefore becomes
\begin{equation}
    \Delta\varepsilon_{\rm tot}= \varepsilon_{\rm tot}^{Q}-\varepsilon_{\rm tot}^{H} = \Delta\varepsilon_{0} +\lambda m_N\left(n_Q-n_H\right),
\end{equation}
where $\Delta\varepsilon_{0}=\varepsilon_Q-\varepsilon_H$ is the energy-density discontinuity in the absence of DM and \(n_H\) is the baryon density on the hadronic phase of the transition. Although the coexistence pressure and baryonic critical chemical potential are unchanged in this limit, the energy-density discontinuity is modified because $n_Q\neq n_H$.

\begin{figure}[t!]
   \centering	
	\includegraphics[width=0.48\textwidth]{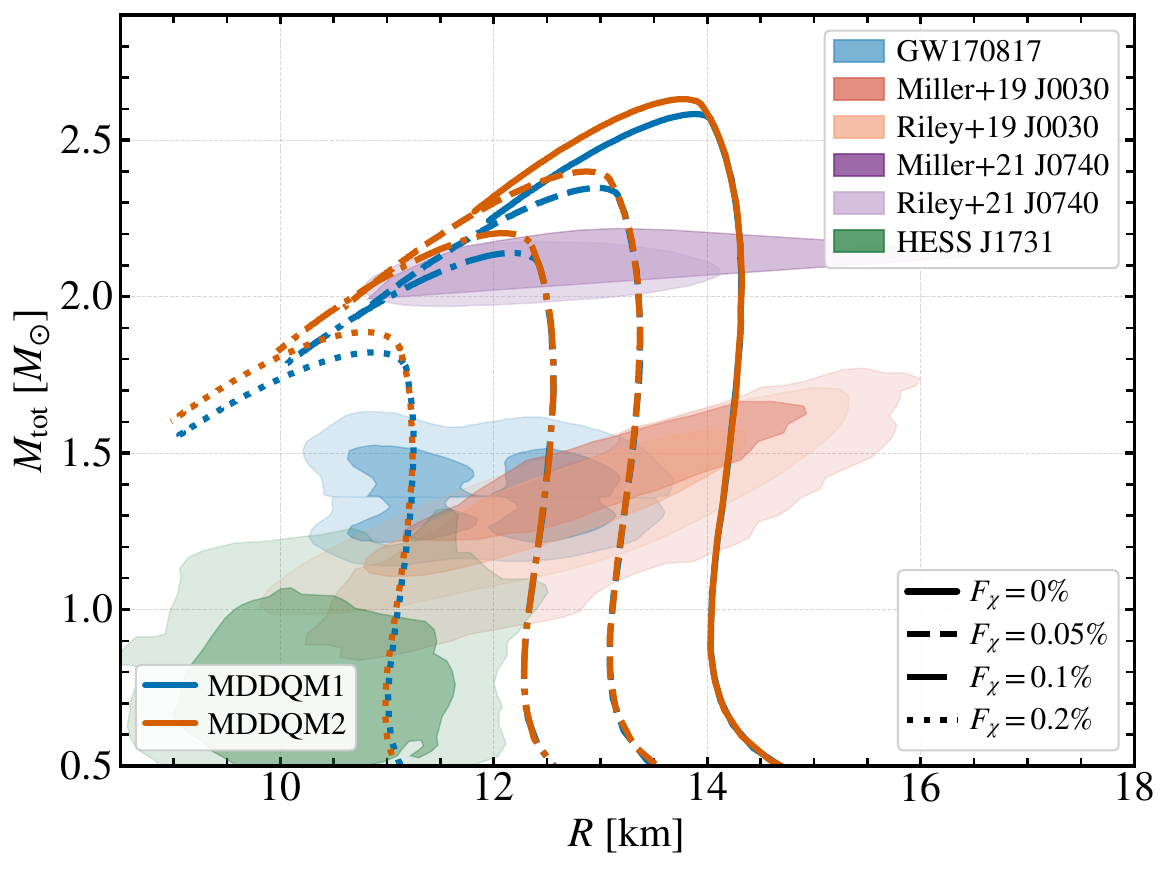}
    	\caption{Gravitational mass versus radius for neutron stars obtained with the NL3$\omega\rho$ hadronic model combined with the two modified DDQM quark-matter parametrizations, MDDQM1 and MDDQM2, for different $F_\chi$. The line colors distinguish the two MDDQM parametrizations, while the line styles denote the DM fraction. The shaded confidence regions show observational constraints from GW170817, HESS~J1731--347, PSR~J0740+6620, and PSR~J0030+0451. For GW170817~\cite{LIGOScientific:2018mvr}, the outer and inner contours correspond to the $90\%$ and $50\%$ credible regions, respectively. For HESS~J1731--347, they correspond to the $95\%$ and $68\%$ credible regions~\cite{2022NatAs...6.1444D}. The PSR~J0740+6620 constraints are shown at the $95\%$ confidence level from the analyses of Miller \textit{et al.}~\cite{Miller:2021qha} and Riley \textit{et al.}~\cite{Riley:2021pdl}, while the PSR~J0030+0451 constraints are shown at the $95\%$ confidence level from the analyses of Miller \textit{et al.}~\cite{Miller:2019cac} and Riley \textit{et al.}~\cite{Riley:2019yda}.}
\label{MR_curve}
\end{figure}
\Cref{MR_curve} shows a suppression of the mass--radius sequences in hybrid MDDQM stars compared to the purely BM case. While both MDDQM1 and MDDQM2 reach $M_{\max}\sim2.6\,M_\odot$ at $F_\chi=0\%$, consistent with the $\sim2\,M_\odot$ constraint from PSR~J0740+6620~\cite{Miller:2021qha,Riley:2021pdl} and the GW170817~\cite{LIGOScientific:2018mvr} and PSR~J0030+0451~\cite{Miller:2019cac, Riley:2019yda} regions, increasing $F_\chi$ reduces $M_{\max}$ to $\sim1.85$--$1.9\,M_\odot$ at $F_\chi=0.2\%$, below the J0740 constraint at 95\% confidence level and toward the HESS~J1731 candidate region~\cite{2022NatAs...6.1444D}. The stronger suppression relative to the pure-BM case 
reflects the combined softening from DM admixture and QM. The transition-induced kink persists with increasing $F_\chi$, while the sequences shift to smaller radii, making $F_\chi\gtrsim0.2\%$ increasingly difficult to reconcile with massive-pulsar constraints.

\subsection{Universal DM scaling}
\begin{figure*}[t!]
   \centering	
	\includegraphics[width=1\textwidth]{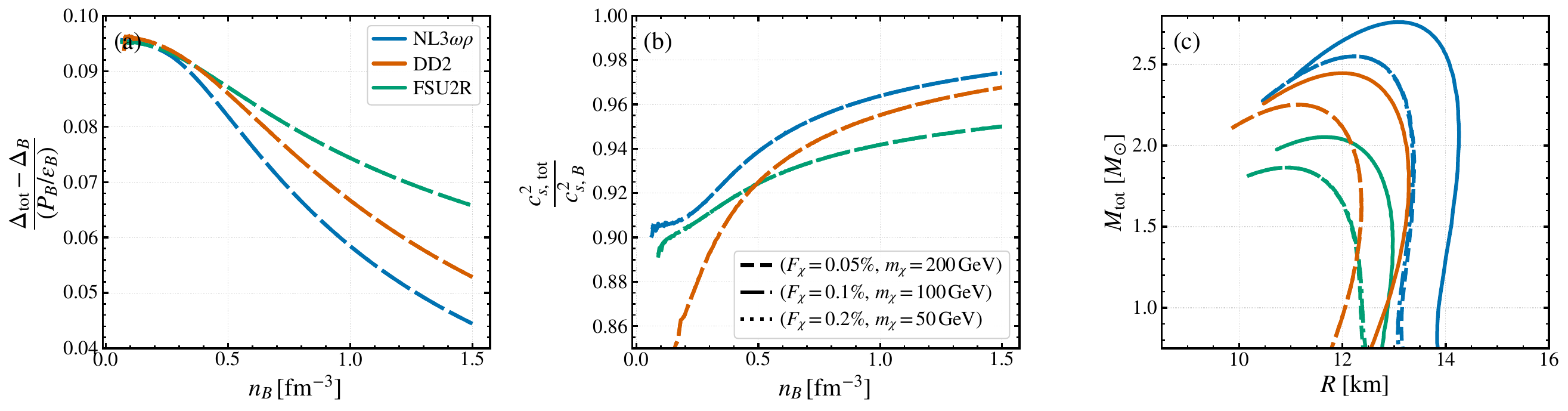}
    	\caption{Numerical test of the universal DM mass-loading scaling for three combinations of $(F_\chi,m_\chi)$ chosen to give approximately the same $\lambda=F_\chi m_\chi/m_N\simeq 0.11$. {Colors distinguish the NL3$\omega\rho$, DD2, and FSU2R EOSs, while line styles denote the different $(F_\chi,m_\chi)$ combinations: $(0.05\%,200\,\mathrm{GeV})$, $(0.1\%,100\,\mathrm{GeV})$, and $(0.2\%,50\,\mathrm{GeV})$}. Panels (a), (b), and (c) show, respectively, the normalized trace-anomaly shift $(\Delta_{\rm tot}-\Delta_B)/(P_B/\varepsilon_B)$, the sound-speed ratio $c_{s,\rm tot}^2/c_{s,B}^2$, and the corresponding mass--radius relations. Curves for different values of $(F_\chi,m_\chi)$ overlap except for the low-mass mass--radius curves. The solid lines in panel (c) indicate configurations with no DM component, $\lambda=0$, compared with curves with the same $\lambda\simeq 0.11$ and different $(F_\chi,m_\chi)$, which show overlap. These demonstrate that, in the heavy-WIMP regime, the leading stellar response is primarily controlled by $\lambda$.  The differences between the curves reflect the underlying BM EOS dependence.}
\label{universal}
\end{figure*}

\Cref{universal} provides a direct numerical test of the proposed $\lambda$-scaling and demonstrates that the universality resides in the dark-sector dependence rather than in the underlying BM EOS. In this numerical test, we use the full EOS, including all interaction terms, to test the validity of the universal scaling and assess the accuracy of the first-order approximations. {We consider the combinations $(0.05\%,200\,\mathrm{GeV})$, $(0.1\%,100\,\mathrm{GeV})$, and $(0.2\%,50\,\mathrm{GeV})$ all giving $\lambda\simeq0.11$, yet they differ substantially in $F_\chi$ and $m_\chi$.} In panel~(a), the normalized trace-anomaly shift follows the same behavior for the three $(F_\chi,m_\chi)$ combinations at fixed $\lambda$, while the separation between NL3$\omega\rho$, DD2, and FSU2R reflects their distinct BM responses. Panel~(b) shows an additional test using the speed of sound propagation ratio for a fixed $\lambda\simeq 0.11$;  their overlapping $c_{s,\rm tot}^2/c_{s,B}^2$ curves show that the sound speed modification is governed by $\lambda$ alone in the heavy-WIMP limit. Panel~(c) shows that this scaling extends to the stellar mass--radius relation, with equal-$\lambda$ sequences remaining closely grouped for each EOS. Thus, the dark-sector parameters exhibit an approximate scaling degeneracy: for a fixed BM EOS and within the heavy, comoving, pressureless regime, the leading stellar response depends on $F_\chi$ and $m_\chi$ through the single mass-loading parameter $\lambda$, while the differences reflect the model dependence of the BM sector. A reduction of the low-median mass star radius means that if, in a given galaxy, the median radius of these stars is smaller than in other galaxies, this could be due to the presence of DM. However, in the case of the pressureless comoving DM we have considered in the present study,  the radius would only give information on the parameter $\lambda$ and not on the DM mass and fraction independently. 

\section{Conclusion}\label{cnc}

We demonstrate that a small DM admixture with NS matter can leave a distinct and quantifiable imprint on the conformal structure of the dense matter without triggering a phase transition of its own. The trace-anomaly decomposition, $\Delta_{\rm tot}=\sum_a w_a\Delta_a\simeq(1-w_\chi)\Delta_B+w_\chi\Delta_\chi$, isolates the underlying mechanism. For the collider-motivated Higgs-portal benchmarks considered, the explicit Higgs, vector-mediator, and contact-interaction contributions are negligible. Instead, the dominant effect arises from the DM rest-mass energy density weighting a nonrelativistic component with $\Delta_\chi\simeq1/3$, reflecting the pressureless scenario. Owing to the mass hierarchy $m_\chi\gg m_N$, even a particle fraction $F_\chi = N_\chi/N_B\sim0.2\%$ can produce a large upward shift of $\Delta_{\rm tot}$ by $\sim0.1$. This enhancement is generic throughout the heavy-WIMP regime ($m_\chi\gtrsim5~\mathrm{GeV}$), whereas $\Delta_\chi$ departs from $1/3$ only for sub-GeV DM, where relativistic effects become important. 

In the pressureless, comoving heavy-WIMP limit defined by the conditions $P_{\rm tot}\simeq P_B$ and $n_\chi=F_\chi n_B$, these results further reduce the leading dark-sector dependence to the single parameter $\lambda=F_\chi m_\chi/m_N$. Numerical calculations, including the microscopic DM--BM interactions, with distinct $(F_\chi,m_\chi)$ pairs at fixed $\lambda$ show overlapping normalized trace-anomaly, sound-speed ratio, and mass--radius responses for a given BM EOS, demonstrating that the stellar response is insensitive to $F_\chi$ and $m_\chi$ separately. This agreement supports the proposed universal scaling with $\lambda$, as the inclusion of the microscopic interactions does not produce any visible deviation from the scaling relations as shown in \cref{universal} within the numerical accuracy of our calculations. The EOS dependence reflects the underlying BM dynamics rather than a breakdown of the dark-sector scaling. The central implication of the universal DM scaling framework is that NS observables constrain the dark-sector mass loading, $\lambda=F_\chi m_\chi/m_N$, rather than $m_\chi$ and $F_\chi$ separately. Consequently, within the regime considered here, NS observations alone cannot disentangle $m_\chi$ from $F_\chi$. Doing so requires independent information on either $m_\chi$ or $F_\chi$, which complementary astrophysical or particle-physics observations could provide. This provides a compact parameterization for incorporating NS observations into searches for nonrelativistic DM and establishes $\lambda$ as the natural quantity for future observational and theoretical studies.

In hybrid stars, DM-induced modifications and genuine deconfinement exhibit qualitatively distinct signatures. While DM leaves the hadron--quark coexistence conditions, characterized by $P_{\rm cr}$ and $\mu_{\rm cr}$, essentially unchanged, a first-order phase transition produces sharp discontinuities in $c_s^2$, $\Delta_{\rm tot}$, and $\Delta_{\rm tot}-\Delta_B$. Thus, WIMP-like DM primarily modifies the energy-density weighting and global stellar structure rather than the phase-coexistence conditions. The combined effects of DM-induced softening and deconfinement substantially increase compactness and suppress the maximum stellar mass, placing $F_\chi=0.2\%$ in increasing tension with the existence of $\sim2\,M_\odot$ NSs. Taken together, the results establish a clear distinction between dark-sector mass loading, which produces a continuous and $\lambda$-controlled modification of the stellar properties, and genuine first-order deconfinement, which introduces discontinuities in the thermodynamic observables. The combined behavior of the trace anomaly, sound speed, and stellar structure provides a unified thermodynamic framework for distinguishing these two mechanisms. WIMP-like DM smooths the density dependence of the trace anomaly and shifts it upward. Consequently, a negative (positive) baryonic trace anomaly is driven toward (away from) the conformal limit, $\Delta=0$.

\begin{acknowledgments}
A.~I. and T.~F.  thank CFisUC for hosting the collaboration.
A. I. acknowledges financial support from the São Paulo State Research Foundation (FAPESP), Grant Nos. 2023/09545-1 and 2025/17347-0. 
T. F. thanks to FAPESP, Grant Nos. 2023/13749-1, 2024/17816-8 and 2025/05312-8, and to the Conselho Nacional de Desenvolvimento 
Cient\'{i}fico e Tecnol\'{o}gico (CNPq), Grant No. 306834/2022-7. This work is a part of the
project Instituto Nacional de  Ci\^{e}ncia e Tecnologia - F\'{\i}sica
Nuclear e Aplica\c{c}\~{o}es  Proc. No. CNPq 408419/2024-5. This work was partially supported by national funds from FCT (Fundação para a Ciência e a Tecnologia, I.P, Portugal) under project UID/04564/2025, identified by DOI 10.54499/UIDB/04564/2025.
\end{acknowledgments}


\appendix
\section{Total energy density and pressure} \label{apendA}
\begin{widetext}
The resulting total energy density, $\varepsilon_{\rm tot}$ and the total pressure $P_{\rm tot}$ are given as:
\begin{align}
\varepsilon_{\rm tot} &=\sum_{i=n,p}\frac{\gamma_B}{2\pi^2}\int_0^{k_{Fi}} dk\,k^2\sqrt{k^2+m_N^{*2}}+\sum_{l=e,\mu}\frac{1}{\pi^2}\int_0^{k_{Fl}} dk\,k^2\sqrt{k^2+m_l^2}
+\frac{\gamma_\chi}{2\pi^2}\int_0^{k_{F}^\chi} dk\,k^2\sqrt{k^2+m_\chi^{*2}}+\frac12m_\sigma^2\sigma_0^2\nonumber\\&+\frac13b(g_\sigma\sigma_0)^3+\frac14c(g_\sigma\sigma_0)^4 +\frac12m_\omega^2\omega_0^2
+\frac{\xi}{8}g_\omega^4\omega_0^4+\frac12m_\rho^2\rho_{03}^2+3\Lambda_\omega g_\omega^2g_\rho^2\omega_0^2\rho_{03}^2 +\frac12m_h^2h^2+\frac12m_X^2X_0^2, \label{engd}
\\
P_{\rm tot}&=\sum_{i=n,p}\frac{\gamma_B}{6\pi^2}\int_0^{k_{Fi}} dk\,\frac{k^4}{\sqrt{k^2+m_N^{*2}}}+\sum_{l=e,\mu}\frac{1}{3\pi^2}\int_0^{k_{Fl}} dk\,\frac{k^4}{\sqrt{k^2+m_l^2}}
+\frac{\gamma_\chi}{6\pi^2}\int_0^{k_{F}^\chi} dk\frac{k^4}{\sqrt{k^2+m_\chi^{*2}}}\nonumber\\&-\frac12m_\sigma^2\sigma_0^2-\frac13b(g_\sigma\sigma_0)^3-\frac14c(g_\sigma\sigma_0)^4 +\frac12m_\omega^2\omega_0^2+\frac{\xi}{24}g_\omega^4\omega_0^4 +\frac12m_\rho^2\rho_{03}^2+\Lambda_\omega g_\omega^2g_\rho^2\omega_0^2\rho_{03}^2-\frac12m_h^2h^2 +\frac12m_X^2X_0^2. \label{pres}
\end{align}
\end{widetext}

\section{DM coupling to the QM} \label{apendB}
The DM sector follows \cref{HP,eq:Vector}, with the couplings generalized from the nucleon to the quark level. The Higgs couples directly to the quark fields through the Standard Model Yukawa interaction \cite{Arcadi:2019lka, Kawamura:2018cpu, Bishara:2015cha},
\begin{equation}
    \mathcal{L}_{qh} =\sum_{q=u,d,s} y_q\,\bar q q\,h, \qquad  y_q=\frac{m_{q0}}{v},
\end{equation}
leading to $ m_q^*=m_{q0}-y_q h$. The DM--Higgs coupling remains unchanged from \cref{HP}. Likewise, the nucleon--mediator interaction in \cref{eq:Vector} is replaced by the quark current interaction
\begin{equation}
    \mathcal{L}_{qX} =-\sum_{q=u,d,s} g_{qX}\,\bar q\gamma^\mu q\,X_\mu .
\end{equation}
In the mean-field approximation, the effective quark chemical potential is $\mu_q^*=\mu_q+\mu_I-g_{qX}X_0$,
where $\mu_I$ denotes the interaction contribution, and
\begin{equation}\label{veom}
    X_0=
    \frac{g_{qX}(n_u+n_d+n_s)+g_{\chi X}n_\chi}
    {m_X^2}.
\end{equation}
The quark phase satisfies beta equilibrium and charge neutrality, with
\begin{equation} \label{qbaden}
    n_Q=\frac{n_u+n_d+n_s}{3},
    \qquad
    n_\chi=F_\chi n_Q,
\end{equation}
where $n_Q$ is the baryon density at the quark phase. So, the same self-consistent DM prescription is maintained across the transition. The thermodynamically consistent MDDQM pressure and energy density associated with the DDQM models are taken from \cite{Issifu:2025byg} and references therein. These quantities are then matched to the hadronic EOS through the Maxwell conditions above to determine $P_{\rm cr}$, $\mu_{\rm cr}$, and the transition densities.

\bibliography{references}

\end{document}